\documentclass{PoS}

\newcommand{\atel}{ATEL}
\newcommand{\apj}{ApJ}

\newcommand{\apjs}{ApJS}
\newcommand{\aj}{AJ}
\newcommand{\aap}{A\&A}
\newcommand{\mnras}{MNRAS}

\newcommand{\pasp}{PASP}

\newcommand{\arXiv}{arXiv}

\newcommand{\integral}{INTEGRAL}
\newcommand{\swift}{\emph{Swift}}
\newcommand{\chandra}{\emph{Chandra}}
\newcommand{\spitzer}{\emph{Spitzer}}
\newcommand{\planck}{\emph{Planck}}
\newcommand{\glimpse}{GLIMPSE}
\newcommand{\mipsgal}{MIPSGAL}

\newcommand{\igr}{IGR\,J17448-3232}
\newcommand{\cxopt}{CXOU\,J174437.3-323222}
\newcommand{\cxoex}{CXOU\,J174453.4-323254}

\newcommand{\arcsec}{$^{\prime\prime}$}
\newcommand{\arcmin}{$^{\prime}$}
\newcommand{\arcdeg}{$^{\circ}$}

\title{IGR J17448-3232 point source: A blazar candidate viewed through the Galactic centre?\thanks{Based on observations collected at the European Organisation for Astronomical Research in the Southern Hemisphere, Chile under ESO program 084.D-0535 (P.I. Chaty)}}
\ShortTitle{IGR J17448-3232: A blazar candidate viewed through the Galactic centre?}

\author{\speaker{Peter A. Curran}$^a$, S.~Chaty$^a$, J.A.~Zurita Heras$^b$, J.A.~Tomsick$^c$, T.J.~Maccarone$^d$ \\
  $^a$AIM, CEA/DSM - CNRS, Irfu/SAP, Centre de Saclay, Bat. 709, FR-91191 Gif-sur-Yvette Cedex, France\\
  $^b$Fran\c{c}ois Arago Centre, APC, Universit\'e Paris Diderot, CNRS/IN2P3, CEA/DSM, Observatoire de Paris, 13 rue Watt, 75205 Paris Cedex 13, France \\
  $^c$Space Sciences Laboratory, 7 Gauss Way, University of California, Berkeley, CA 94720-7450, USA\\
  $^d$School of Physics and Astronomy, University of Southampton, Southampton, Hampshire, SO17\,1BJ, UK\\
        E-mail: \email{peter.curran@cea.fr}
}

\abstract{
  The error region of the \integral\ source, \igr, contains an X-ray
  point source at the edge of a $\sim3$\arcmin\ radius extended X-ray
  source.  It has been suggested that the extended emission is a young
  supernovae remnant (SNR) while the point source may be an isolated
  neutron star, associated with the SNR, that received a kick when the
  supernova occurred. We identify the infrared counterpart of the
  X-ray point source, visible from 2.2 $\mu$m to 24 $\mu$m, and place
  limits on the flux at longer wavelengths by comparison with radio
  catalogues. Multi-wavelength spectral modeling shows that the data
  are consistent with a reddened and absorbed single power law over
  five orders of magnitude in frequency. This implies non-thermal,
  possibly synchrotron emission that renders the previous
  identification of this source as a possible pulsar, and its
  association to the SNR, unlikely; we instead propose that the
  emission may be due to a blazar viewed through the plane of the
  Galaxy.
}

\FullConference{The Extreme and Variable High Energy Sky\\
		 September 19-23, 2011\\
		 Chia Laguna (Cagliari), Italy}

\begin{document}

\section{Introduction}\label{section:introduction}

We observed the field of \igr\ as part of our observing campaign at
ESO's New Technology Telescope (NTT). The purpose of this program is
to detect near-infrared (nIR) counterparts of \integral\ sources in
order to identify or confirm their nature via Spectral Energy
Distributions (SEDs) and/or spectra
\cite{Chaty2008:A&A484,Rahoui2008:A&A484}.  The main focus of the
program is to investigate the numerous intrinsically obscured, high
mass X-ray binaries (HMXBs) observed by \integral\ and so far we have
been able to identify 13 of these as either Supergiant (5 cases) or Be
(8 cases) systems via spectral analysis [Coleiro et al. {\it in
  preparation}]. As many of the \integral\ sources are not previously
detected in the nIR or optical, there is often confusion regarding
their true nature.  Hence, along with the HMXBs identified in our
sample, we observed and later classified a number of sources as low
mass X-ray binaries (LMXBs) \cite{Curran2011:A&A533}, pulsar wind
nebulae (PWNe) \cite{Curran2011:A&A534} and, in the case of \igr\
which we summarise here, a blazar \cite{Curran2011:MNRAS}.

\begin{figure} 
  \centering 
  \resizebox{\hsize}{!}{\includegraphics[angle=-90]{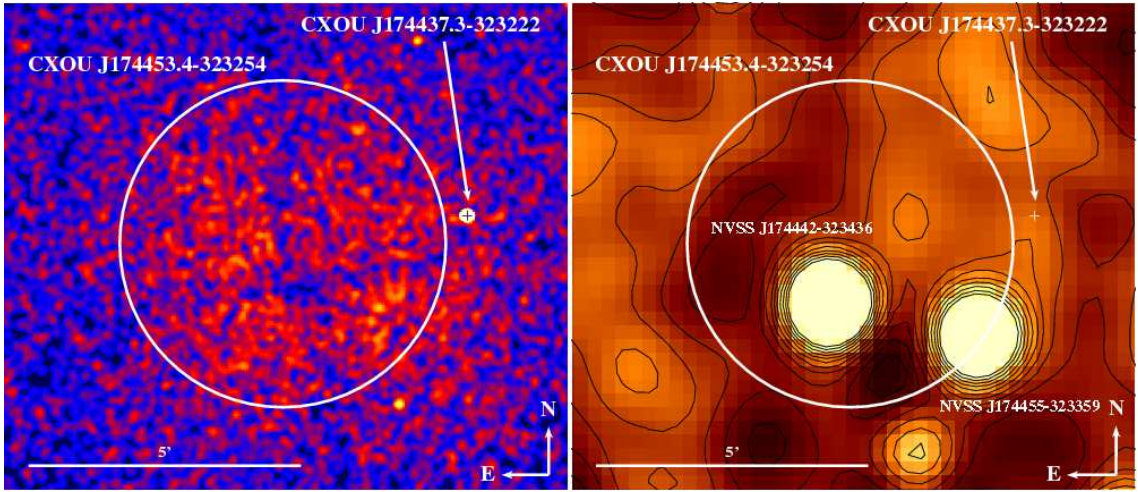} } 
  \caption{ The field of \igr\ ($\sim 10$\arcmin$\times10$\arcmin) as
    detected in the X-rays (0.3-10 keV) by \chandra\ ({\it left}) and
    in radio (1.4\,GHz) by the NRAO VLA Sky
    Survey ({\it right}).  The $\sim3$\arcmin\
    radial extent of the extended emission (white circle) is clearly
    visible in the X-ray image, as is the blazar candidate point
    source. In the radio image two nearby NVSS/MGPS-2 radio sources
    are visible but there is no emission consistent with the central
    position of either  the extended or the point source. }
 \label{fig:field} 
\end{figure}

The X-ray source, \igr\ was initially discovered by INTEGRAL and
published in the Third and subsequently Fourth IBIS/ISGRI Soft
Gamma-Ray Survey Catalog \cite{Bird2007:ApJS170,Bird2010:ApJS186} at
slightly different, though consistent, positions and fluxes.  In an
attempt to refine the position, the \emph{Swift} X-ray telescope (XRT)
observed the field and a point source at the edge of the original
\integral\ error circle, as well as possible diffuse emission, were
identified \cite{Landi2007:ATel.1323}.  The position of the point
source was further refined by a 4.7\,ks \chandra\ observation of the
field \cite{Tomsick2009:ApJ701}. In addition to detecting the point
source, \cxopt, those authors confirmed an extended source, $\sim7$
times brighter, at the \integral\ position (\cxoex, see
Figure\,\ref{fig:field}, {\it left}).  At the XRT point source
position a $R \sim 15.5$ USNO-B1.0 \cite{Monet2003:AJ125} and
$K=9.100$ 2MASS \cite{Skrutskie2006:AJ.131} source was noted
\cite{Landi2007:ATel.1323}, however, the sub-arcsecond accuracy of the
\chandra\ position eliminates this source as a possible counterpart;
though the X-ray point source is at the edge of the 2MASS point spread
function.
Based on analysis of the \chandra\ spectrum it was
proposed \cite{Tomsick2009:ApJ701} that the extended emission
originated from a young supernova remnant (SNR), though no evidence of
a pulsar wind nebula (PWN) was detected. A tentative association of
the point source with the SNR was also suggested, the authors
hypothesizing that it may be an isolated neutron star which received a
kick when the supernova occurred.

\section{The search for a counterpart}\label{Observations}

We collated data of \igr\ from various sources including published
catalogs, archived images and our own ESO-NTT observations, in an
attempt to identify multi-wavelength counterparts and investigate the
nature of the source.
In the high energy range ($\gtrsim 10$\,keV) a source is documented in
the Palermo Swift-BAT hard X-ray catalogue (15--150\,keV;
\cite{Cusumano2010:A&A524}), though given the resolution of the BAT
instrument this is likely an unresolved measurement of both the
extended and point source. No source, consistent with either position,
is found in the \emph{Fermi} LAT First Source Catalog (100\,MeV --
100\,GeV; \cite{Abdo2010:ApJ188}) and no such source has been made
public by \emph{Fermi} GBM (10\,keV -- 30\,MeV) or HESS (100\,GeV --
100\,TeV; \cite{Chaves2009:arXiv0907}).
At the other end of the spectrum, the second epoch Molonglo Galactic
Plane Survey (MGPS-2) compact source catalogue
\cite{Murphy2007:MNRAS382}, which details observations at 843\,MHz,
documents two nearby sources which are also included in the NRAO VLA
Sky Survey (1.4\,GHz; NVSS; \cite{Condon1998:AJ115}).  The extension
of these sources overlap with the extended X-ray emission
but neither are consistent with the point-like X-ray source
(Figure\,\ref{fig:field}, {\it right}). From a visual inspection of
the NVSS and MGPS-2 images available of the field, there is no obvious
sign of any excess emission above background levels at the point
source position so we use the flux density of a nearby dim object as a
measure of the 1.4\,GHz upper limit in the field.  No nearby sources
are found in any of the 9 bands (30\,GHz -- 856\,GHz) of the all-sky,
\planck\ Early Release Compact Source Catalog
\cite{Planck2011:arXiv1101} and while flux density limits for the
region are not well quantified, they are on the order of 1\,Jy
\cite{Planck2011:arXiv1101}.

\begin{figure}
  \centering 
  \resizebox{\hsize}{!}{\includegraphics[angle=-0]{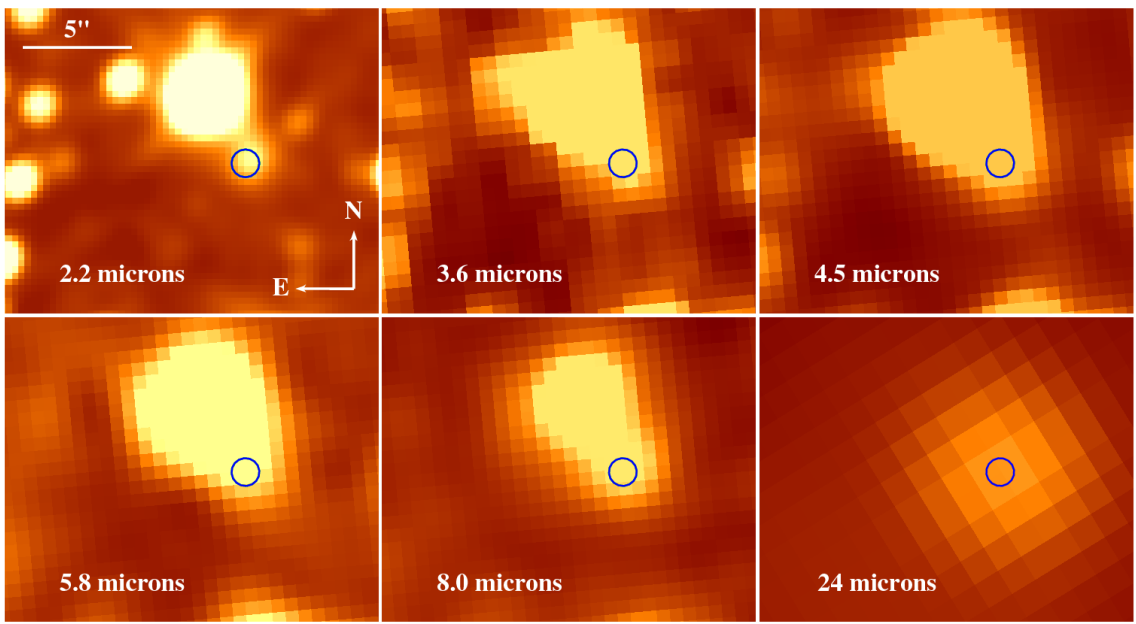} } 
\caption{
The infrared images of the \chandra\ point source position (0.64\arcsec\
90\% error circle marked) show a clearly detected counterpart at all wavelengths. The
images are from our ESO-NTT $K_{S}$ band observations (2.2$\mu$m),
archival \spitzer\ IRAC images (3.6-8.0$\mu$m) and the \spitzer\ MIPSGAL
survey (24$\mu$m). 
}
 \label{fig:2-24um} 
\end{figure}

Moving to higher energies, we obtained a $K_{S}$ band (2.2 $\mu$m)
image of the field with the SofI instrument on the 3.58m ESO-New
Technology Telescope (NTT).  We also utilised data from the
\emph{Spitzer Space Telescope}'s \glimpse\ \cite{Benjamin2003:PASP115}
and \mipsgal\ \cite{Carey2009:PASP121} surveys with entries at 3.6, 4.5,
5.8, 8.0 and 24 $\mu$m. While there is no sign of any diffuse
emission corresponding to the extended X-ray emission in any of
images, a resolved counterpart is clearly detected at the point source position
(Figure \ref{fig:2-24um}).
The probability of a chance superposition down to the observed
magnitude of the source in the $K_{S}$ band is $\sim3$\%, or down to
the limiting magnitude of the field is $\sim10$\%, while the
probabilities are lower (1-4\%) at the more sparsely populated longer wavelengths.
We were also able to obtain an upper limit from \swift's
Ultraviolet/Optical Telescope (UVOT) $uw2$ band ($\sim 0.2 \mu$m); the
lack of a detection is not surprising, given the high value of
Galactic extinction in the direction of the Galactic centre and the
blueness of the $uw2$ filter.

\section{Multi-wavelength Spectral Energy Distribution}\label{section:SED}

\begin{figure}
  \centering 
  \resizebox{\hsize}{!}{\includegraphics[angle=-90]{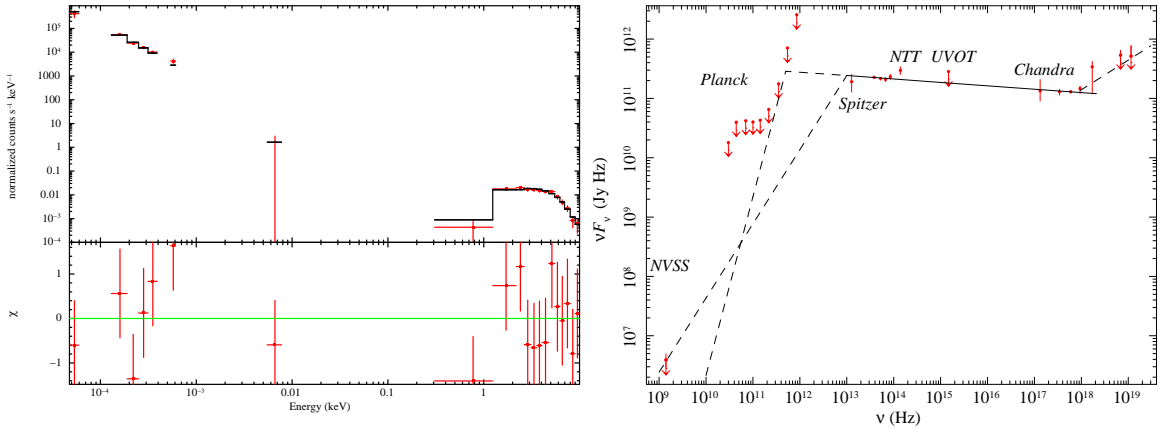} } 
  \caption{{\it Left:} Absorbed and reddened power law {\tt XSPEC} fit
    ($\alpha = -1.057$) of the \spitzer, NTT-SofI, \swift-UVOT\ and
    \chandra\ data. Also included is a secondary, phenomenological
    power law component to eliminate an excess at energies $\gtrsim
    5$keV.  {\it Right:} Unabsorbed/dereddened SED for the point
    source, showing the power law fit ($\alpha = -1.057$) and two
    possible low frequency spectral slopes ($\alpha = 0.25$, $\alpha =
    2.0$).  Also included are the high energy upper limits defined
      by the unresolved \integral\ and BAT
      fluxes which limit the steepest
      possible high energy slope to be $\alpha \sim -0.5$, assuming a
      break at $\sim10^{18}$\,Hz.
}
 \label{fig:spectra} 
\end{figure}

The infrared, \swift\ UVOT and \chandra\ X-ray data were fit within
{\tt XSPEC} by a single power law (Figure\,\ref{fig:spectra} {\it
  Left}) , affected by interstellar extinction and absorption.
However, this model has a clear excess of emission at energies
$>5$\,keV that could not be accounted for by any possible
pile-up. Including an additional, purely phenomenological, power law
component produces a marginal improvement of the fit with power law
and extinction/absorption parameters similar to those of the initial
fit.  The best fit (Figure\,\ref{fig:spectra}) gives a broadband
spectral index, $\alpha$ ($F_{\nu} \propto \nu^{\alpha}$) from
infrared to X-ray of $\alpha = -1.057 \pm 0.015$ ($1\sigma$
confidence), as well as a secondary spectral index to account for the
excess emission $>5$\,keV which is poorly constrained. 
If the observed, unresolved \integral\ and BAT fluxes (at
$\sim6\times10^{18}$\,Hz and $\sim1\times10^{19}$\,Hz respectively)
are taken as upper limits on the contribution of the point source at
these energies, we can place an approximate limit on the steepest
possible high energy slope of $\alpha \sim -0.5$, assuming a break at
$\sim10^{18}$\,Hz.

The optical extinction is fit as $E_{(B-V)} = 0.84^{+0.3}_{-0.15}$ and
the equivalent hydrogen column density as $N_{\mathrm{H}} = 2.69 \pm
0.25 \times 10^{22}$\,cm$^{-2}$, marginally greater than the Galactic value of
$N_{{\rm H\,Galactic}} = 0.67 \times 10^{22}$\,cm$^{-2}$
 \cite{kalberla2005:A&A440}.  These should be treated with caution due
to the small range over which they are calculated and, in the case of
extinction, the lack of sensitivity at the observed  wavelengths.

The \planck\ limits imply that the spectrum breaks at a frequency
$\sim 10^{12}$\,Hz, i.e., between the \planck\ and
\spitzer\ frequencies, though the exact value will depend on the
sharpness of the break. The flattest possible spectral index, $\alpha$
at low frequencies is $\approx 0.25$, while the upper limits are also
consistent with a steepest possible (physical) spectral slope of
$\alpha=2$.

\section{A blazar candidate viewed through the Galactic centre?}\label{section:discussion}

The IR to X-ray SED displays a single power law with no evidence for
any thermal emission at any frequency in the observed bands -- only a
suggestion of excess emission above $5$\,keV -- and the  spectral
index of this non-thermal emission corresponds to the expected
spectral slope of synchrotron emission from accelerated electrons 
(though this is only one possible interpretation of the data).
As non-contemporaneous X-ray flux measurements from \swift-XRT and
\chandra\ are consistent, we can assume that the source is relatively
steady, though we cannot rule out variability.  The single power
law SED derived from non-contemporaneous archive/catalog values also
supports this assumption. 
This persistent, single power law emission might be expected from a
number of astronomical sources such as AGN, persistent X-ray binaries,
microquasars or magnetars but in most of these cases there would be a
measurable thermal component, of which there is no evidence of
here. Power law emission would also be expected from a SNR or PWN but
the extended nature of these sources should be observable in either
the X-ray or IR/optical, which is not the case. As mentioned above,
this source has been suggested to be a pulsar but a pulsar would not
be expected to be so radio dim relative to its brightness in the nIR
regime.


One solution that can explain the emission is that the source is a
blazar, an AGN with its jet pointing directly at us (e.g.,
\cite{Ghisellini2011:arXiv1104}). Blazars are persistent sources,
though it should be noted that they have been observed to undergo
flares and high energy rapid variability as well as long term
evolution, to which the observations detailed here would not be
sensitive.  In this scenario the emission in a given regime is
dominated by synchrotron or inverse Compton (IC) emission from the jet
which, because of its angle towards us, is much brighter than the
thermal and other emission associated with the AGN.  Even though the
source is close to the Galactic centre ($\sim 3.6$\arcdeg), the
Galactic column density is relatively low in that direction, allowing
the emission to be visible through the Galaxy. The possible excess
column density implied by the X-ray spectra suggests excess absorption
which may be explained by absorption close to the source.  The
measured IR and X-ray fluxes of this
source
and spectral slope are all broadly consistent with those of blazars in
general \cite{Fossati1998:MNRAS.299}.  As blazars are associated with
radio galaxies, we would generally expect radio emission from their
jets but in this case we do not, though the implied radio limit is
consistent with a low frequency synchrotron spectral slope
(Figure\,\ref{fig:spectra} {\it Right}) and the detected radio fluxes
in other blazars\cite{Fossati1998:MNRAS.299}.

In this framework the apparent excess of emission at energies
$>5$\,keV ($\sim10^{18}$\,Hz) which we model, phenomenologically, with
a second power law is due to the expected inverse Compton emission
from blazars.  However, we can only place a loose constraint on this
excess component from the \chandra\ spectrum and the upper limits
implied by the unresolved (from the extended source) emission in the
higher energy bands (i.e., \swift-BAT, \integral).

Tests of this hypothesis require an optical or IR spectrum with which
to confirm the extragalactic nature of the source, though blazars, if
of the BL Lac class of radio galaxy, are expected to have no or only
very weak lines. Hence, it may be the absence of lines, such as those
that would be expected from Galactic sources, that will add weight to
the blazar argument. However, the main test is a broader-band SED
spanning from radio to optical and into the high energy X-rays, which
should display the double peaks (synchrotron and inverse Compton), or
at least slopes, and confirm the absence of a thermal component.
Approved X-ray observations of the source will also allow us to test
whether any possible variability is consistent with a blazar.
Additionally, a high level of linear polarisation might be expected if
the source is a blazar, due to the synchrotron emission.

\section{Conclusions}\label{section:conclusions}

On the basis of positional coincidence and common spectral slope, we
have identified a new infrared counterpart to the X-ray point source
in the field of \igr, visible from 2.2 to 24\,$\mu$m.
Multi-wavelength spectral modelling shows that the data are consistent
with a reddened and absorbed single power law over five orders of
magnitude in frequency. This implies non-thermal, possibly synchrotron
emission that we propose may be due to a blazar viewed through the
plane of the Galaxy. If this source is confirmed as a blazar, at $l,b =
356.81, -1.70$ degrees, while not being the first in the Galactic plane, it
will be the first identified so close to the Galactic centre.
If the point source is a blazar it is possible that the extended
emission, instead of being a SNR, is in fact a Galaxy cluster although
it is also possible that the two sources of emission  are not
associated; approved X-ray  observations will allow us to
confirm the nature of the extended emission.

\acknowledgments
We thank the referee for their constructive comments. 
This work was supported by the Centre National d'Etudes Spatiales
(CNES) and is based on observations obtained with MINE: the
Multi-wavelength INTEGRAL NEtwork.

\end{document}